\documentclass[prl,aps,amssymb,amsmath,superscriptaddress,twocolumn]{revtex4}
\usepackage{amsmath}
\usepackage{amssymb}
\usepackage{epsfig}
\usepackage{amscd}
\usepackage{graphicx,psfrag,xspace}
\usepackage{float}
\usepackage[normalem]{ulem}

\begin{document}

\title{Discrete Feynman-Kac formulas for branching random walks}
\author{A. Zoia}
\email{andrea.zoia@cea.fr}
\affiliation{CEA/Saclay, DEN/DANS/DM2S/SERMA/LTSD, 91191 Gif-sur-Yvette, France}
%\affiliation{Commissariat \`a l'Energie Atomique et aux Energies Alternatives, Direction de l'Energie Nucl\'eaire, D\'epartement de Mod\'elisation des Syst\`emes et Structures, Service d'Etudes des R\'eacteurs et de Math\'ematiques Appliqu\'ees, CEA/Saclay, 91191 Gif-sur-Yvette, France}
\author{E. Dumonteil}
\affiliation{CEA/Saclay, DEN/DANS/DM2S/SERMA/LTSD, 91191 Gif-sur-Yvette, France}
\author{A. Mazzolo}
\affiliation{CEA/Saclay, DEN/DANS/DM2S/SERMA/LTSD, 91191 Gif-sur-Yvette, France}

\begin{abstract}
Branching random walks are key to the description of several physical and biological systems, such as neutron multiplication, genetics and population dynamics. For a broad class of such processes, in this Letter we derive the discrete Feynman-Kac equations for the probability and the moments of the number of visits $n_V$ of the walker to a given region $V$ in the phase space. Feynman-Kac formulas for the residence times of Markovian processes are recovered in the diffusion limit.
\end{abstract}
\maketitle

\begin{figure}[b]
 \centerline{\epsfclipon \epsfxsize=9.0cm
\epsfbox{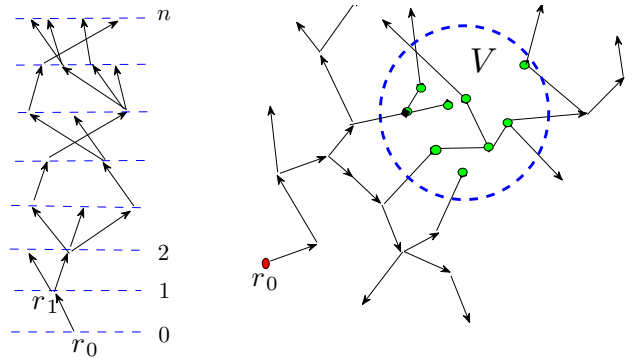} }
\caption{(Color online). Branching random walks. Left: the evolution of a path as a function of the generation number. Right: a path starting from ${\mathbf r}_0$ and performing a number $n_V$ of visits to the region $V$.}
 \label{fig1}
\end{figure}

Consider the walk of a particle starting from a point-source and performing random displacements. At given times, the particle disappears by giving rise to $k=0,1,2,\cdots$ descendants with probability $p_k$, the case $p_0$ corresponding to an absorption. Each descendant behaves exactly as the mother particle, the overall path resulting in a branched structure, as shown in Fig.~\ref{fig1}. Branching random walks lie at the heart of physical and biological modeling~\cite{harris, athreya, jagers}, and are key to the description of neutron multiplication and nucleon cascades~\cite{pazsit}, disordered systems~\cite{korolev}, evolution of biological populations~\cite{lawson}, diffusion of reproducing bacteria~\cite{barret, golding}, and mutation-propagation of genes~\cite{sawyer}, just to name a few. Detailed accounts of the huge research efforts devoted to this subject, ranging from the pioneering work by Galton and Watson on the extinction probability to the more recent developments, can be found, e.g., in the monographs~\cite{harris, athreya, pazsit}. In particular, even the simplest of these models, namely, a Brownian particle that at exponentially distributed times gives rise to two descendants, turns out to be highly nontrivial, and is still widely investigated, in view of its both practical and conceptual interest, especially in relation with such issues as front propagation and extreme statistics~\cite{derrida, derrida_barrier, spohn, vansaarloos, majumdar_extreme}.

A central question for random walks is to assess the sojourn time $t_V$ of the stochastic paths in a given portion $V$ of the phase space~\cite{condamin_benichou, condamin, majumdar_occupation, grebenkov, agmon_lett}: for branching Brownian motion, this subject was first explored in connection with problems in mathematical genetics~\cite{sawyer, iscoe, cox}. The evolution of a number of physical and biological systems, such as neutrons or populations, is most often described in terms of discrete generations, so that one is quite naturally led to consider the number $n_V$ of visits to the domain $V$, as illustrated in Fig.~\ref{fig1}. When $V$ is large with respect to the typical displacement size, we can safely assume $n_V \propto t_V$, which roughly amounts to assuming that the underlying walk can be  approximated by a Brownian motion. However, it is well known that this simple proportionality breaks down when $n_V$ is small, and application of the diffusion approximation might therefore lead to inaccurate results~\cite{blanco, zdm_prl, zdm_pre}.

In this Letter, we derive a discrete Feynman-Kac approach to characterize the distribution $P_n(n_V|{\mathbf r}_0)$ and the associated moments for a branching random walk that is observed up to the $n$-th generation, for arbitrary displacements, offspring distributions, and geometries. This provides a general framework for dealing with a broad class of discrete-time random walks, which are ubiquitous in physics and biology.

{\em Feynman-Kac formulas.} A single walker is initially isotropically emitted from ${\mathbf r}_0$, and undergoes a sequence of displacements and collisions. We assume that the displacements from ${\mathbf r'}$ to ${\mathbf r}$, between any two collisions, are equally distributed, and obey the probability density $T({\mathbf r'};{\mathbf r})$. At each collision, the incident particle disappears, and $k$ particles are emitted isotropically with probability $p_k$. For the sake of simplicity, we furthermore assume that $p_k$ is spatially homogeneous. We formally define the number $n_V$ of visits to $V$ as $n_V(n)=\sum_{i} V({\mathbf r}_i)$, where $V({\mathbf r}_i)$ is the marker function of the region $V$, which takes the value $1$ when the point ${\mathbf r}_i \in V$, and vanishes elsewhere, and the sum is extended to all the points visited by the source particle and its descendants up to the $n$-th generation. We assume here that the source point ${\mathbf r}_0$ is not counted. Clearly, $n_V$ is a stochastic variable, depending on the realizations of the underlying process, and on the initial condition ${\mathbf r}_0$. The behavior of its distribution, $P_n(n_V|{\mathbf r}_0)$, is most easily described in terms of the associated generating function
\begin{equation}
F_n(u|{\mathbf r}_0)= \langle u^{-n_V}\rangle_n({\mathbf r}_0) = \sum^{+\infty}_{n_V=0}P_n(n_V|{\mathbf r}_0)u^{-n_V}.
\end{equation}
This approach was first proposed by Kac, based on Feynman path integrals, for continuous-time Markov processes (see, e.g.,~\cite{kac, majumdar_review}), and recently extended to non-Markovian walks~\cite{turgeman}. The quantity $F_n(u|{\mathbf r}_0)$ is a stochastic functional (over the branching paths) that can be thought of as the discrete Laplace transform of $P_n(n_V|{\mathbf r}_0)$. In order to derive an evolution equation for $F_n(u|{\mathbf r}_0)$, we closely follow here the argument in~\cite{zdm_pre}. We initially consider trajectories starting with a single particle entering its first collision at ${\mathbf r}_1$. At ${\mathbf r}_1$, $k$ particles are created, with probability $p_k$. Random flights are Markovian at collision points, which allows splitting each of the $k$ subsequent trajectories into a first jump, from ${\mathbf r}_1 $ to ${\mathbf r}_1 + \Delta_k$ (the displacement $\Delta_k$ obeying the jump length density $T$), and then a branching path from ${\mathbf r}_1 + \Delta_k$ to the positions held at the $n$-th generation. If $k=0$, the trajectory ends at ${\mathbf r}_1 $ and there will be no further events contributing to $n_V$. Hence, we have
\begin{eqnarray}
\tilde{F}_{n+1}(u|{\mathbf r}_1)=p_0u^{-V({\mathbf r}_1)}+p_1u^{-V({\mathbf r}_1)}\langle \tilde{F}_n(u|{\mathbf r}_1 + \Delta_1)\rangle+\nonumber \\
+p_2 u^{-V({\mathbf r}_1)}\langle \tilde{F}_n(u|{\mathbf r}_1 + \Delta_{2}')\tilde{F}_n(u|{\mathbf r}_1 + \Delta_{2}'')\rangle + \cdots ,
\label{equation_F_one}
\end{eqnarray}
where expectation is taken with respect to the random displacements $\Delta$, and $u^{-V({\mathbf r}_1)}$ can be singled out because it is not stochastic. The tilde is used to recall that we are considering trajectories starting with a single particle entering the first collision at ${\mathbf r}_1$. The terms in Eq.~\eqref{equation_F_one} can be understood as follows: the probability that $k$ identical and indistinguishable particles (born at ${\mathbf r}_1$) give rise to $n_V$ collisions in $V$ is given by the convolution product that the first makes $n_{1}$ collisions, the second $n_{2}$, ..., and the $k$-th $n_V-n_{1}-n_{2}-\cdots$. In the transformed space, this convolution becomes a simple product of generating functions. If we assume that the offspring particles are independent, the expectation of the products in Eq.~\eqref{equation_F_one} becomes the product of the expectations. We make then use of the discrete Dynkin's formula
\begin{equation}
\langle f({\mathbf r}_1+ \Delta)\rangle = \int T^*({\mathbf r}'; {\mathbf r}_1) f({\mathbf r}')d{\mathbf r}',
\end{equation}
where $f$ is any sufficiently well behaved function of a stochastic process, and $T^*({\mathbf r}'; {\mathbf r})$ is the adjoint kernel  associated to $T({\mathbf r}'; {\mathbf r})$~\cite{zdm_pre}. Intuitively, $T^*$ displaces the walker backward in time. We therefore obtain the discrete Feynman-Kac equation
\begin{equation}
\tilde{F}_{n+1}(u|{\mathbf r}_1)
=u^{-V({\mathbf r}_1)} G\left[ \int T^*({\mathbf r}'; {\mathbf r}_1)\tilde{F}_n(u|{\mathbf r}')d{\mathbf r}'\right] ,
\label{equation_F}
\end{equation}
where $G[z]=\sum_{k=0}^{+\infty}p_k z^k$ is the generating function of $p_k$. Equation~\eqref{equation_F} plays a central role, in that it relates the generating function $F_n$ of the number of visits $n_V$ to the generating function $G$ of the offspring number $k$. Finally, by observing that the first collision coordinates ${\mathbf r}_1$ obey the probability density $T({\mathbf r}_0;{\mathbf r}_1)$, it follows
\begin{equation}
F_n(u|{\mathbf r}_0)=\int \tilde{F}_n(u|{\mathbf r}_1) T({\mathbf r}_0;{\mathbf r}_1)d{\mathbf r}_1,
\label{feynman_x0}
\end{equation}
together with the initial condition $\tilde{F}_1(u|{\mathbf r}_1)=u^{-V({\mathbf r}_1)}$. Knowledge of $F_{n}(u|{\mathbf r}_0)$ allows determining $P_n(n_V|{\mathbf r}_0)$: indeed, $F_{n}(u|{\mathbf r}_0)$ is a polynomial in the variable $u$, the coefficient of each power $u^{-i}$ being $P_n(n_V=i|{\mathbf r}_0)$.

{\em Moments formulas.} A complementary tool for characterizing the distribution $P_n(n_V|{\mathbf r}_0)$ is provided by the analysis of its moments. By construction, $\tilde{F}_n(u|{\mathbf r}_1)$ is the (rising) factorial moment generating function for trajectories entering their first collision at ${\mathbf r}_1$, which implies
\begin{equation}
\langle \tilde{n}_V^{(m)} \rangle_n({\mathbf r}_1) = (-1)^m \frac{\partial^m}{\partial u^m} \tilde{F}_n ( u|{\mathbf r}_1 ) \vert_{u=1},
\label{recursion_mom_disc}
\end{equation}
$x^{(k)}=x(x+1)...(x+k-1)$ being the rising factorial~\cite{pitman_book}. The tilde is again used to recall that the moments refer to trajectories starting from the first collision at ${\mathbf r}_1$. Combining Eqs.~\eqref{equation_F} and~\eqref{recursion_mom_disc} and using Fa\`{a} di Bruno's formula~\cite{pitman_book} for the $m$-th derivative of the composite function $G[\tilde{F}_n({\mathbf r}_1)]$ yields the recursion property for $m \ge 1$
\begin{eqnarray}
\langle \tilde{n}_V^{(m)} \rangle_{n+1}({\mathbf r}_1)=mV({\mathbf r}_1)\langle \tilde{n}_V^{(m-1)} \rangle_{n+1}({\mathbf r}_1)+\nonumber \\
+\sum_{j=1}^{m} \nu_j {\cal B}_{m,j} \left[\langle n_V^{(1)} \rangle_n({\mathbf r}_1) ,\cdots, \langle n_V^{(m-j+1)} \rangle_n({\mathbf r}_1)\right],
\label{recursion_F_n}
\end{eqnarray}
where ${\cal B}_{m,j} \left[z_1,\cdots,z_{m-j+1}\right]$ are the Bell's polynomials~\cite{pitman_book}, $\nu_j=\langle k(k-1)...(k-j+1)\rangle$ are the falling factorial moments of the offspring number, and we have used $\int T^*({\mathbf r}'; {\mathbf r}_1)\langle \tilde{n}_V^{(q)} \rangle_n({\mathbf r}')d{\mathbf r}'=\langle n_V^{(q)} \rangle_n({\mathbf r}_1)$. Bell polynomials~\footnote{ ${\cal B}_{1,1} [z_1]=z_1;  {\cal B}_{2,1} [z_1,z_2]=z_2, {\cal B}_{2,2} [z_1,z_2]=z^2_1;...$.} commonly appear in connection with the combinatorics of branched structures~\cite{pitman_book}: this might give a hint about their role in Eq.~\eqref{recursion_F_n}, which relates the moments $\langle \tilde{n}_V^{(m)} \rangle_{n}$ of the visit number  to the moments $\nu_j$ of the descendant number. Observe that $\langle \tilde{n}_V^{(1)} \rangle_{n}$ depends only on $\nu_1$, $\langle \tilde{n}_V^{(2)} \rangle_{n}$ on $\nu_1$ and $\nu_2$, and so on. The recurrence is initiated with the conditions $\langle \tilde{n}_V^{(0)} \rangle_n({\mathbf r}_1)=1$, and $\langle \tilde{n}_V^{(m)} \rangle_1({\mathbf r}_1)=m!V({\mathbf r}_1)$. Finally, the factorial moments $\langle n_V^{(m)} \rangle_n({\mathbf r}_0)$ for particles emitted at ${\mathbf r}_0$ are obtained from
\begin{equation}
\langle n_V^{(m)} \rangle_n({\mathbf r}_0) = \int \langle \tilde{n}_V^{(m)} \rangle_n({\mathbf r}_1) T({\mathbf r}_0 ; {\mathbf r}_1)d{\mathbf r}_1.
\label{moments_T_r0}
\end{equation}

\begin{figure}[t]
 \centerline{\epsfclipon \epsfxsize=9.0cm
\epsfbox{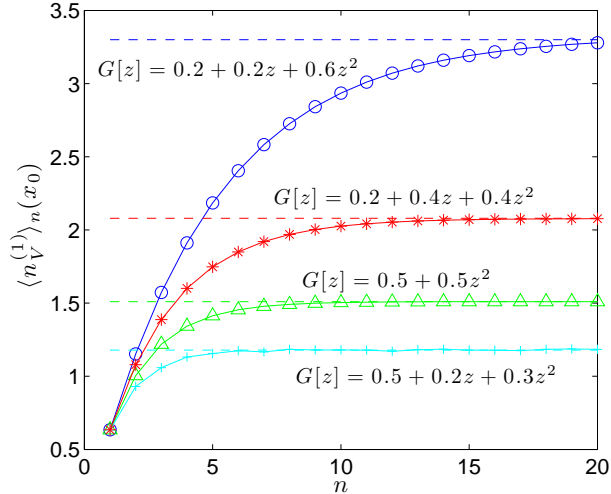} }
\caption{(Color online). Exponential flights on the interval $[-R,R]$: first factorial moment of $n_V$ as a function of $n$. Here $x_0=0$, $\sigma=1$ and $R=1$. Blue circles: $\nu=1.4$ ($R_c \simeq 1.59$); red stars: $\nu=1.2$ ($R_c \simeq 2.57$); green triangles: $\nu=1$; cyan crosses: $\nu=0.8$. Dashed lines: the stationary moments.}
 \label{fig2}
\end{figure}

{\em Stationary behavior.} When trajectories are observed up to $n\to +\infty$, we define $\langle n_V^{(m)} \rangle=\lim_{n\to +\infty}\langle n_V^{(m)} \rangle_n$: it turns out that the asymptotic moments $\langle n_V^{(m)} \rangle({\mathbf r}_0)$ are related to the stationary distribution of the particles~\cite{zdm_prl}, provided that the limit exists. To see this, we introduce the collision density $\Psi({\mathbf r}|{\mathbf r}_0)$ such that $\Psi({\mathbf r}|{\mathbf r}_0) d{\mathbf r}$ is the particle number at equilibrium in a volume of size $d{\mathbf r}$ around ${\mathbf r}$, for a single particle emitted at ${\mathbf r}_0$~\cite{zdm_prl, zdm_pre}. It can be shown~\cite{zdm_pre_operator} that the collision density satisfies the Boltzmann-like stationary integral transport equation
\begin{equation}
\Psi({\mathbf r}|{\mathbf r}_0)=\nu \int T({\mathbf r}'; {\mathbf r})\Psi({\mathbf r}'|{\mathbf r}_0)d{\mathbf r}'+
T({\mathbf r}_0; {\mathbf r}),
\label{integral_transport_equations}
\end{equation}
where $\nu=\nu_1$ is the average number of descendants per generation. Branching processes in the absence of spatial constraints are said to be subcritical for $\nu < 1$, supercritical for $\nu >1$, and critical for $\nu= 1$, respectively~\cite{harris, athreya, pazsit}. Equation~\eqref{integral_transport_equations} can be understood as follows: the equilibrium particle density at ${\mathbf r}$ for a source emitting at ${\mathbf r}_0$ is given by the sum of all contributions having a collision at ${\mathbf r}'$, being multiplied and then transported to ${\mathbf r}$, plus the contribution of the particles emitted from the source and never collided up to entering ${\mathbf r}$. Now, from the properties of the Bell's polynomials, the $m$-th order term can be singled out of the sum in Eq.~\eqref{recursion_F_n}, so that we can rewrite the stationary moment equation for $\langle \tilde{n}_V^{(m)} \rangle$ as
\begin{eqnarray}
\langle \tilde{n}_V^{(m)} \rangle({\mathbf r}_1)-\nu\int T^*({\mathbf r}'; {\mathbf r}_1)\langle \tilde{n}_V^{(m)} \rangle({\mathbf r}')d{\mathbf r}'  = g_m({\mathbf r}_1),
\label{recursion_F_asy}
\end{eqnarray}
where $g_m({\mathbf r}_1)= mV({\mathbf r}_1)\langle \tilde{n}_V^{(m-1)} \rangle({\mathbf r}_1)+b_m({\mathbf r}_1)$ acts as a source term, and
\begin{eqnarray}
%b_m({\mathbf r}_1)=\sum_{j=2}^{m}\langle k_{(j)}\rangle {\cal B}_{m,j} \left[\tilde{N}^1({\mathbf r}_1) ,\cdots, \tilde{N}^{m-j+1}({\mathbf r}_1)\right],
b_m({\mathbf r}_1)=\sum_{j=2}^{m}\nu_j{\cal B}_{m,j} \left[\langle n_V^{(1)} \rangle({\mathbf r}_1) ,\cdots, \langle n_V^{(m-j+1)} \rangle({\mathbf r}_1)\right]
\label{eq_g}
\end{eqnarray}
represents the contributions from $p_{k \ge 2}$, and vanishes when $p_0+p_1=1$. The integral equation~\eqref{recursion_F_asy} for $\langle \tilde{n}_V^{(m)} \rangle({\mathbf r}_1)$ can be explicitly solved (see~\cite{zdm_pre}), and combined with~\eqref{moments_T_r0} yields
\begin{equation}
\langle n_V^{(m)} \rangle({\mathbf r}_0) = \int\Psi({\mathbf r}'|{\mathbf r}_0) g_m({\mathbf r}') d{\mathbf r}',
\label{stat_moments}
\end{equation}
starting with $g_1({\mathbf r}')=V({\mathbf r}')$, which provides the desired relation between the stationary moments and the collision density. When $p_0+p_1=1$ one recovers the stationary moment formula in~\cite{zdm_prl}.

\begin{figure}[t]
 \centerline{\epsfclipon \epsfxsize=9.0cm
\epsfbox{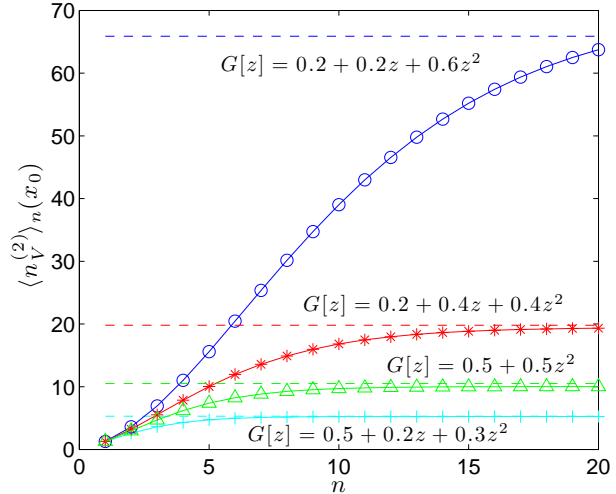} }
\caption{(Color online). Exponential flights on the interval $[-R,R]$: second factorial moment of $n_V$ as a function of $n$. Here $x_0=0$, $\sigma=1$ and $R=1$. Blue circles: $\nu=1.4$; red stars: $\nu=1.2$; green triangles: $\nu=1$; cyan crosses: $\nu=0.8$. Dashed lines: the stationary moments.}
 \label{fig3}
\end{figure}

{\em An example.} We illustrate this formalism on exponential flights, a stochastic process that is widely adopted to model the propagation of neutrons and chemical or biological species, among others (see, e.g.,~\cite{weiss, zdm_prl, zdm_pre}). In one dimension, the displacement kernel reads $T(x';x)=\sigma^{-1}e^{-|x-x'|/\sigma}$, where $\sigma$ is some typical length scale. We assume that the source is $x_0 \in V$. To fix the ideas, we take $V$ as the interval $[-R,R]$, the particles being lost upon crossing the outer boundaries. A natural question is whether the total number of visits will eventually explose or level off to an asymptotic value under the combined effects of the branching mechanism and the spatial leakages. To our best knowledge, the full distribution $P_n(n_V|{\mathbf r}_0)$ for this example is unfortunately not known. Actually, Eq.~\eqref{equation_F} can seldom be inverted, and the nonlinear integral equation resulting from taking the limit $\tilde{F} = \lim_{n\to \infty} \tilde{F}_{n} $ hardly admits explicit solutions. Nonetheless, the moment Eqs.~\eqref{recursion_F_n} and~\eqref{stat_moments} are amenable to exact formulas (at least for simple geometries and displacement kernels, such as in this case) and therefore greatly help in the analysis of branching random walks. Indeed, the essential features of a stochastic system are often captured by the first few moments. The resulting expressions are fairly cumbersome and will therefore not be reported here (hints on how to perform the calculations can be found, e.g., in~\cite{zdm_pre_operator}). Instead, we display the first (Fig.~\ref{fig2}) and second factorial moment (Fig.~\ref{fig3}) as a function of the generation number $n$ for some descendant number probability $p_k$. Analytical results have been verified by Monte Carlo simulation. When $\nu \le 1$ the moments of $n_V$ converge to an asymptotic value; when $\nu >1 $, the moments may converge or diverge depending on whether the leakages from the boundaries compensate the growth of the particle number. This intuitively depends on the existence of a stationary solution $\Psi(x|x_0 )$: we can then define a critical radius as the smallest $R_c$ such that for $R>R_c$ the average number of visits to $V$ diverges when $n\to +\infty$. This quantity can be computed explicitly once $\langle n_V^{(1)} \rangle({\mathbf r}_0)$ is known, and for $\nu>1$ reads $R_c=\sigma \csc^{-1}(\sqrt{\nu})/\sqrt{\nu-1}$, independent of $x_0$.

{\em Diffusion limit.} We conclude by examining the scaling limit of the discrete Feynman-Kac equations, which is obtained when $n_V$ is large, and at the same time the typical jump length $\epsilon$ as well as the net average displacement $\mu$ are vanishing small. We set $t_V=n_V dt$ and $t=n dt$, where $dt$ is some small time scale related to $\epsilon$ by the usual diffusion scaling $\epsilon^2 = 2D dt$ and $\mu=vdt$, $D$ playing the role of a diffusion coefficient, and $v$ of a velocity. By properly taking the limit of large $n_V$ and vanishing $dt$, $t_V$ converges to the residence time in $V$. It is natural to set $p_k=\lambda_k dt$, the quantity $\lambda_k$ being a rate per unit of $dt$. Observe that when $\epsilon$ and $\mu$ are small for any displacement kernel we have the Taylor expansion $\int  T({\mathbf r}'; {\mathbf r})f({\mathbf r}')d{\mathbf r}' \simeq f( {\mathbf r})-\mu \partial_{{\mathbf r}}f( {\mathbf r}) +\frac{1}{2}\epsilon^2\partial^2_{{\mathbf r}}f( {\mathbf r})$, and similarly $\int  T^*({\mathbf r}'; {\mathbf r}_0)f({\mathbf r}')d{\mathbf r}' \simeq f( {\mathbf r}_0)+\mu \partial_{{\mathbf r}_0}f( {\mathbf r}_0) + \frac{1}{2}\epsilon^2\partial^2_{{\mathbf r}_0}f( {\mathbf r}_0)$. It is expedient to introduce the quantity $Q_t(u|{\mathbf r}_0)=F_t(e^{u}|{\mathbf r}_0)$, which is the moment generating function of $t_V$, i.e.,
\begin{equation}
\langle t_V^{m} \rangle_t({\mathbf r}_0) = (-1)^m \frac{\partial^m}{\partial u^m} Q_t(u|{\mathbf r}_0) \vert_{u=0},
\label{recursion_diffusion}
\end{equation}
when trajectories are observed up to time $t$. Under the previous hypotheses, combining Eqs.~\eqref{equation_F} and~\eqref{feynman_x0} and passing to the limit $dt \to 0$ yields
\begin{eqnarray}
\frac{\partial Q_{t}}{\partial t}={\cal L}_{{\mathbf r}_0}^*Q_{t}-uV({\mathbf r}_0)Q_{t}+\lambda G\left[Q_{t}\right].
\label{eq_feynman_diffusion_dt}
\end{eqnarray}
where ${\cal L}_{{\mathbf r}_0}^*= D\partial^2_{{\mathbf r}_0}+v\partial_{{\mathbf r}_0}-\lambda$, and $\lambda=\sum_{k} \lambda_k$. Finally, from Eq.~\eqref{recursion_diffusion} stems the recursion property
\begin{eqnarray}
\frac{\partial \langle t_V^{m} \rangle_t({\mathbf r}_0)}{\partial t} = {\cal L}_{{\mathbf r}_0}^*\langle t_V^{m} \rangle_t({\mathbf r}_0) + mV({\mathbf r}_0)\langle t_V^{m-1} \rangle_t({\mathbf r}_0)+\nonumber \\ 
+\lambda \sum_{j=1}^{m}\nu_j {\cal B}_{m,j} \left[\langle t_V^{1} \rangle_t({\mathbf r}_0),\cdots, \langle t_V^{m-j+1} \rangle_t({\mathbf r}_0) \right].
\end{eqnarray}
As a particular case, when $G[z]=0.5+0.5z^2$ and $v=0$ we recover the results in~\cite{cox} for critical binary branching Brownian motion without drift.

{\em Perspectives.} The Feynman-Kac fomalism proposed in this Letter could be further generalized to take into account spatial dependences ($T({\mathbf r}'; {\mathbf r})$ and/or $p_k$ may vary as a function of ${\mathbf r}$) and anisotropies. A nice application of these results would be to infer the distribution $p_k$ on the basis of the number of countings $n_V$ recorded at a detector.

The authors express their gratitude to Drs.~S.~N.~Majumdar and A.~Rosso for insightful discussions.

\end{document}